\documentclass{aa}
\usepackage{graphicx}
\usepackage{psfig}
\usepackage{natbib}
\bibpunct{(}{)}{;}{a}{}{,}

\begin{document}

\def\spose#1{\hbox to 0pt{#1\hss}}
\def\lta{\mathrel{\spose{\lower 3pt\hbox{$\mathchar"218$}}
     \raise 2.0pt\hbox{$\mathchar"13C$}}}
\def\gta{\mathrel{\spose{\lower 3pt\hbox{$\mathchar"218$}}
     \raise 2.0pt\hbox{$\mathchar"13E$}}}
\def\Msun{{\rm M}_\odot}
\def\msun{{\rm M}_\odot}
\def\Rsun{{\rm R}_\odot}
\def\Lsun{{\rm L}_\odot}
\def\half{{1\over2}}
\def\RL{R_{\rm L}}
\def\zs{\zeta_{s}}
\def\zR{\zeta_{\rm R}}
\def\dJJ{{\dot J\over J}}
\def\dMM{{\dot M_2\over M_2}}
\def\tKH{t_{\rm KH}}
\def\eck#1{\left\lbrack #1 \right\rbrack}
\def\rund#1{\left( #1 \right)}
\def\wave#1{\left\lbrace #1 \right\rbrace}
\def\dd{{\rm d}}
\def\new#1{{#1}}

\title{A transient high-coherence oscillation in 4U 1820-30}

\titlerunning{A transient oscillation in 4U 1820-30}

\author{T. Belloni\inst{1}, I. Parolin\inst{1}
    \and
    P. Casella\inst{2,3}
}

\offprints{T. Belloni}

\institute{INAF -- Osservatorio Astronomico di Brera,
        Via E. Bianchi 46, I-23807 Merate, Italy
   \and
        INAF -- Osservatorio Astronomico di Roma,
        Via di Frascati 33, I-00040 Monte Porzio Catone, Italy
	\and
	Dipartimento di Fisica, Universit\`a degli Studi ``Roma Tre'',
	Via della Vasca Navale 84, I-00146 Roma, Italy
}

\date{Received 26 February 2004; accepted 11 May 2004}

\abstract{We re-analyzed two Rossi X-Ray Timing Explorer archival
observations of the atoll source 4U 1820-30 in order to investigate 
the detailed time-frequency properties of the source during the
intervals when a $\sim$7 Hz QPO was detected by Wijnands
et al. (1999, ApJ, 512, L39). We find that in both observations,
in addition to a QPO signal lasting a couple of minutes as previously
reported, 
there is a much narrower transient oscillation with a life time
of only a few seconds. Within this time, the oscillation is
consistent with being coherent. Its integrated fractional rms is around
10\% and its frequency 7.3 Hz and 5.7 Hz in the two observations.
We discuss the possible association of this QPO with other oscillations
known both in Neutron-Star and Black-Hole systems, concentrating on 
the similarities with the narrow 5-7 Hz oscillations observed at high flux
in Black-Hole Candidates.

}

\maketitle

\section{Introduction}

Since its launch in 1995, the
Rossi X-ray Timing Explorer (RXTE) has provided a wealth of new
information on the timing properties of
accreting X-ray binaries (see van der Klis 2004 for a 
review). In Neutron-Star Low-Mass X-ray Binaries (LMXB), 
oscillations at frequencies from $<$1 Hz to more than 1 kHz are now known,
although there is no 
general consensus yet on their origin. 
The observed phenomenology of these Quasi-Periodic Oscillations (QPO)
and of noise
components in LMXB systems as observed by RXTE is complex. In particular, in 
addition to the high-frequency QPOs (with frequencies between a few hundred
and $\sim$1200 Hz), two types of low-frequency oscillations are observed. 
One with frequencies that vary between 16 and 70 Hz, associated to 
a band-limited noise component (called Horizontal-Branch Oscillation, HBO, 
in bright Z sources), and one with lower
frequencies (4-7 Hz) appearing at high accretion rate in Z sources (called
Normal Branch Oscillation, or NBO). 
In most 
Black-Hole Candidates (BHCs), low-frequency oscillations with frequencies
varying between 0.1 and 30 Hz are known to be present
in at least in some states (see van der Klis
2004). These QPOs are always associated with a band-limited noise component
and their frequency is believed to be positively correlated with mass accretion
rate (see e.g Sobczak et al. 2000). They are called ``type-C'' QPOs
although often they are referred to as `1-10 Hz QPOs'.
A strong connection between the HBO and type-C QPOs described above
was found by Wijnands \& van der Klis (1999), Psaltis, Belloni \&
van der Klis (2000), and Belloni, Psaltis \& van der Klis (2002), who showed
evidence of large scale correlations between their frequencies and the
characteristic frequencies of other aperiodic components. These correlations
include both Neutron-Star and Black-Hole systems. One of them
has been since 2002 extended to White-Dwarf systems
(see Mauche 2002, Warner et al. 2003 and references therein).

\begin{figure}[h] \resizebox{\hsize}{!}{\includegraphics{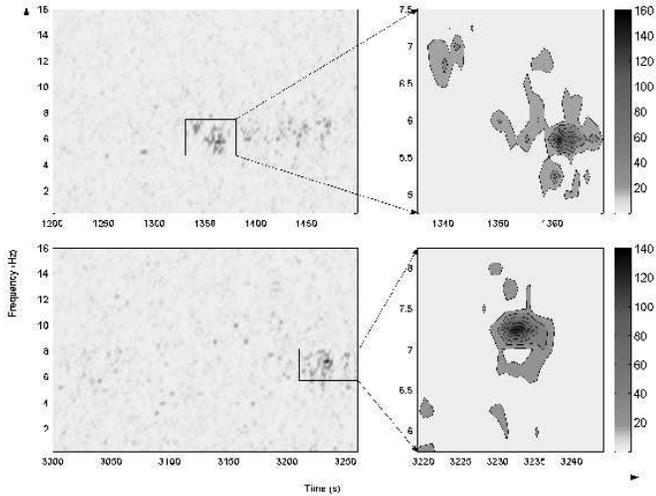}}
\caption{Left panels: grays-cale representation of the spectrograms
for the relevant sections of the 
two observations analyzed here. Right panels: enlarged
sections corresponding to the regions where QPO signal is evident.
The gray-scale bars on the right show the level of Leahy power.
Notice that consecutive 4-s PDS in these images are shifted by only 1
second.
} \label{figure1}
\end{figure}

Two additional classes of low-frequency 
QPOs in BHC have been identified (see Wijnands et al. 1999b, Homan et al. 2001,
Remillard et al. 2002, Nespoli et al. 2003, Casella et al. 2004). The first
class, 
called ``type A'' is characterized by being very broad (with a quality factor 
 $Q$, defined as the centroid frequency divided by the FWHM of the peak in the
 PDS, of less than 3), The second class, called ``type B'', is much narrower
($Q \geq 10$), appears at frequencies around 6 Hz, and is associated to 
a weak power-law noise component. This type-B oscillation shows erratic
variability of its centroid frequency on time scales of $\sim$10 s  and
was observed to 
appear and disappear rapidly (see Nespoli et al. 2003, Casella et al. 2004).
Its frequency is only seen in the rather narrow 5-8 Hz range across a number
of sources. 
The time/phase lag properties of these A-B-C oscillations (and their higher
harmonics when present) are also well
determined and different across the three classes. 

Wijnands et al. (1999a, hereafter W99) 
reported the discovery of a $\sim$7 Hz QPO in the
LMXB 4U 1820-30, an Atoll source. The QPO was transient, lasting only 160
s, the corresponding power spectral peak was
narrow (FWHM$\sim$0.5 Hz), and showed erratic shifts in frequency. It was
found only when the source was at the highest luminosity in the analyzed
data. 
In a note added in proof, W99 mentioned that a similar oscillation
 was also present in another dataset which had just recently become public. 
These authors found an association with BHC QPOs unlikely and
concentrated their attention to the similarities to NBOs. 
In order to examine this oscillation more closely and compare it with 
recent results on BHCs, we reanalized the 4U 1820-30 data where this
oscillation was reported by W99 applying the
same procedures
used by Nespoli et al. (2003) and Casella et al. (2004).

\begin{figure}[h] \resizebox{\hsize}{!}{\includegraphics{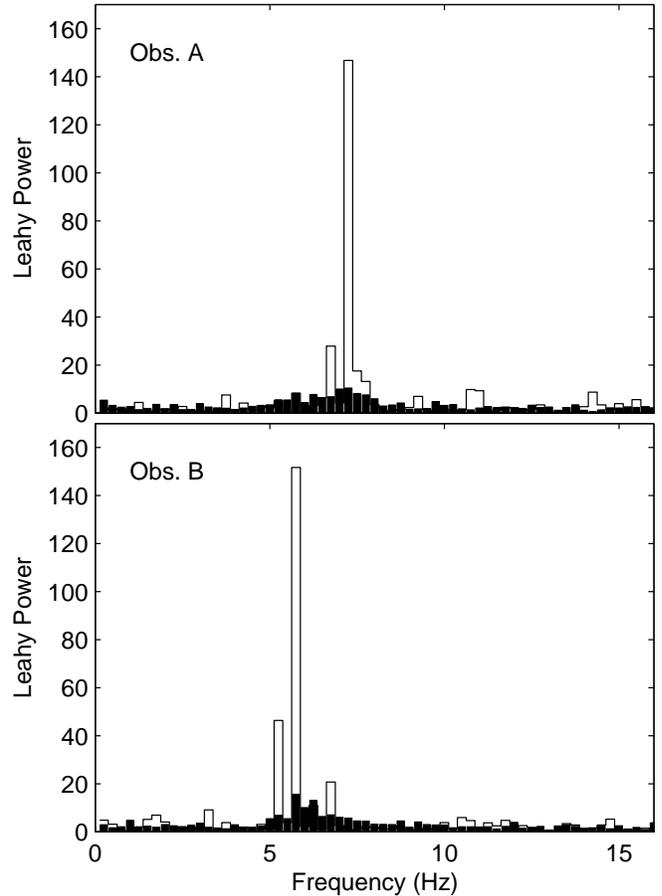}}
\caption{PDS corresponding to the peak power in the spectrograms of
Fig. 1 (empty bars) and average PDS of the preceding and
following four PDS (filled bars).} \label{figure2}
\end{figure}

\section{Data analysis}

We analyzed two RXTE/PCA observations of 4U 1820-30 from the RXTE
public archive (see Table 1). The first observation (observation A) is one
RXTE orbit from the observation showing the sharp QPO (see W99). For the second
observation (observation B), we produced spectrograms for 
all observations presented by Zhang
et al. (1998) and found the only RXTE orbit mentioned by W99 where
the sharp QPO is visible.

\begin{table}
\begin{center}
\caption{\small PCA observations of 4U 1820-30 analyzed in this
work}
\begin{tabular}{lccc}
\hline
\hline
Obs.   &  Start Date & Start Time & Exp. (s)\\
A      &  1996-May-4 & 17:06      & 3512    \\
B      &  1997-jul-16& 5:12       & 3640    \\

\hline

\hline
\end{tabular}
\end{center}
\end{table}

In order to investigate the time-frequency behaviour of the time 
series for each
observation, we produced a spectrogram, defined (in its continuous
version) as

\begin{equation}
S_x (t,\nu) =\left|{\int_{-\infty}^{+\infty} x(u) h^\ast (u-t)
  e^{-i2\pi \nu u} du}\right|^2 $$
\end{equation}

(see Nespoli et al. 2003), where $x(t)$ is the signal accumulated
over PCA channels 0-35 (corresponding approximately to the energy
range 2-13 keV) with a time resolution of 1/32 seconds,
and $h(t)$ is a window function (in our case a
4-seconds long boxcar window). The spectrogram resolution was
chosen to be 1 second, which means that the resulting spectrogram
consists of power density spectra from a sliding window with time
step 1 second. The spectrograms for the interesting intervals of
the two observations 
(normalized according to Leahy et al. (1983) are shown
in gray scale in the left panels of Fig. 1. The right panels show
an enlarged contour-plot view of the sections where oscillations are visible.
The gray-scale bars indicate that high values of power are present in this
region.

Figure 1 suggests that the signal is composed of two parts: a broad
excess and a sharp peak with a very short duration. We produced
new spectrograms without overlap between consecutive segments (i.e. with 
time step 4 seconds) and searched for the highest
peak in each of the two observations. A maximum power 
of 146.8 in Obs. A and 151.7 in Obs B was found.
In Fig. 2 we show for each observation the PDS corresponding to these
maxima and the average of the four observations before and after.
It is clear that each PDS with the maximum power is different and
much more peaked than the surrounding ones. Side peaks are visible in both
PDS: as their distance is a multiple of 0.5 Hz, they are related to effects of
the 4-s window.
In order to check whether
these strong peaks could be simple statistical deviations from the
broad QPO peak in the surrounding spectra, we applied the procedures
described by van der Klis (1989, section 3.4.2). 
From each of the two observations, we averaged 
the PDS from four intervals before and four after the
one containing the sharp peak in order to estimate the local shape of the
broad component. We then fitted them with a Lorentzian model
(see Table 1), and computed the best-fit power 
$<P_b>$ at the frequency
of the sharp peak. If $P_s$ is the power of the sharp peak of which we
want to test the significance, if the underlying signal is the same, 
the quantity $2P_s / <P_b>$ is distributed as a $\chi^2$ with two degrees
of freedom. The one-trial significance of the sharp peaks for the two
observations is then $6.1\times 10^{-8}$ and $9.9\times 10^{-7}$ respectively.
W99 detect the broad oscillations for $\sim$160 s, while in our second
observation significant signal is present for about 100 seconds. Allowing 
for these numbers of trials, the detected sharp peaks are very significant.

\begin{table*}
\begin{center}
\caption{\small Lorentzian parameters of the detected QPOs 
}
\begin{tabular}{lcccccc}
\hline
\hline
Obs. &  \multicolumn{3}{c}{Narrow peak}                      & \multicolumn{3}{c}{Broad peak}                   \\
\hline
     &  $\nu_0$ (Hz)   & $\Delta$ (Hz) & rms                 & $\nu_0$ (Hz) & $\Delta$ (Hz) & rms               \\
\hline
A    &   7.28$\pm$0.03 & 0.16$\pm$0.03 & 10.64$\pm$0.61\%    & 7.02$\pm$0.09& 1.97$\pm$0.31 & 7.73$\pm$0.52\%   \\
B    &   5.73$\pm$0.05 & 0.14$\pm$0.05 &  9.21$\pm$1.14\%    & 6.01$\pm$0.05& 1.20$\pm$0.16 & 7.01$\pm$0.38\%   \\

\hline
\hline
\end{tabular}
\end{center}
\end{table*}

We fitted the features shown in Fig. 2 with a Lorentzian model, obtaining the 
parameters reported in Table 1. As one can see, the FWHM $\Delta$ of the
narrow peaks is smaller than or compatible with 
the frequency resolution of 0.25 Hz of the
PDS. This indicates that the width of these peaks can be entirely due
to windowing effects, and therefore the oscillation is consistent with
being coherent within those 4 seconds. Notice that there is little evidence
for a red-noise component, as one can also see from Fig. 2 in 
Wijnands et al. (1999). The short duration of our intervals prevents a
sensitive measurement or upper limit to the red noise variability,
but W99 estimate a 3.4\% upper limit to the 0.1-1 Hz low-frequency noise
in Obs. A, consistent with our results.

In order to investigate the shape of the oscillation, we folded the 
4-s light curves at the best-fit period from Table 2. The folded
light curves can be seen in Fig. 3. Their shape is close to sinusoidal. 
Fits
with a sinusoidal function show that a weak second harmonic component
{\bf (fractional rms $\sim$2\%)} might be
present in Obs. A, corresponding to the weak peak observable in Fig. 2. The
residuals from a sinusoidal shape for Obs. B are more complex than a simple
harmonic peak: {\bf a good fit can be obtained only with two additional harmonic
components, with a barely significant second harmonic (rms$\sim$1\%) and
a stronger third harmonic (rms$\sim$2.4\%) responsible for the bump observable
in Fig. 3.}
Finally, we produced cross spectra for both observations using the
energy bands 2.0-6.9 keV and 6.9-18.1 keV for Obs. A, and 
2.0-6.5 keV and 6.5-18.1 keV for Obs. B (due to different accumulation
modes). From the cross spectra, we estimated the time lag of the
hard curve with respect to the soft one to be +2.0$\pm$0.8 ms for Obs. A
and -5.6$\pm$4.2 ms for Obs. B (positive lag correspond to the hard
curve lagging the soft one). Clearly, the short duration prevents
a meaningful determination of the lags. Notice that for Obs. A. we obtain
results consistent with those of W99 when we use a larger time interval,
therefore including also the broad QPO peak.

\begin{figure}[h] \resizebox{\hsize}{!}{\includegraphics{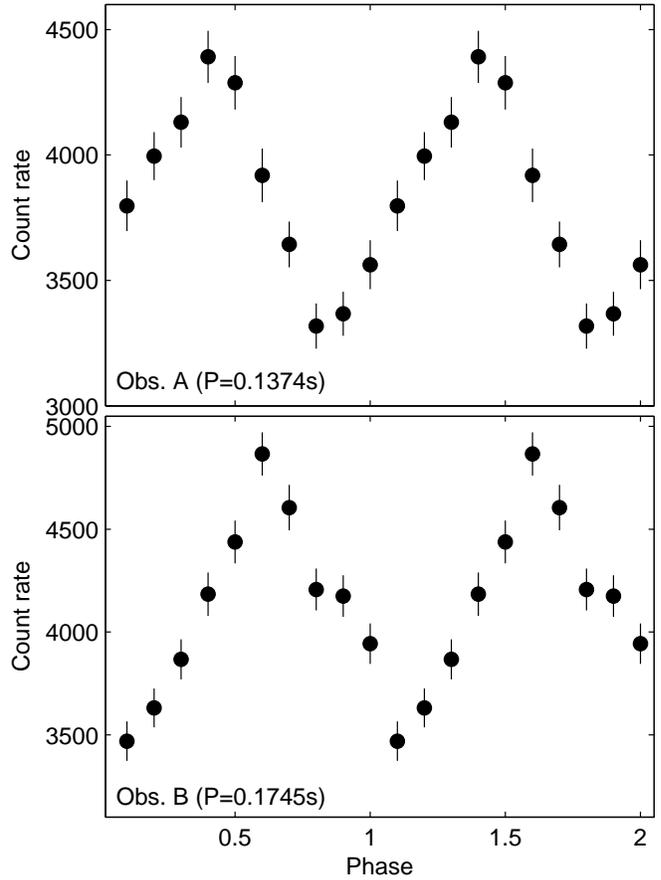}}
\caption{Folded light curves for the two intervals with the
high-coherence oscillation. Two cycles are shown for clarity.} \label{folding}
\end{figure}

We looked for the presence of high-frequency oscillations in correspondence
to the short intervals where the 5-7 Hz QPO was detected. 
In Observation A the highest available frequency for such a search at
low energies (where most of the photons are detected) is 256 Hz, while
for Obs. B we explored frequencies up to 2048 Hz. We did not find evidence
for a high-frequency oscillation, although clearly the short duration
of the QPO interval limits rather severely the sensitivity of such a 
search. We note, however, that
Zhang et al. (1998) reported that kHz observations were 
detected in their observations only when the source was soft, independent
of the source state. Our 7 Hz QPO appears when the source is in the
hardest region of the color-color diagram see W99).

\section{Discussion}

As discussed by W99, the two observations analyzed in this work
show a signal at a frequency of a few Hertz, lasting only 
a short time. However, the oscillation reported by
W99 corresponds to the black histogram in Fig. 2, due to the
limited time resolution of their analysis. We found that in both
observations, in addition to the broader signal of W99, a much
narrower peak is present in the spectrogram for a shorter
period of time. These oscillations are peculiar, in that they 
last only about four seconds and within that time are highly coherent. A
4-s interval corresponds to $\sim$29 and $\sim$23 cycles for
Obs. A and B respectively. Also, the oscillations are strong, with 
about a 10\% fractional rms.

QPOs with these characteristics have not been observed before in 
bright Neutron-Star LMXBs. As discussed by W99, in both cases they 
appear when 4U 1820-30 is in the uppermost region of its
banana branch, corresponding to the highest observed luminosity
(about 5$\times 10^{37}$erg/s, see W99). 
However, within one single observation in this state, the QPO appears
in a section of the light curve that displays no other peculiarities
such as e.g. an anomalous flux or color with respect to the surrounding
intervals.

W99 discard a possible association with the type-C QPOs observed in BHCs,
which at the time were the only ones known, but of course do not discuss the
more similar type-B QPOs.

Therefore, they tentatively associate the broad oscillation with the 5-20 Hz
QPOs observed in Z sources. However, the properties we detected in 
the narrow QPOs are not seen in 
these oscillations from
Z sources, although no high time resolution searches have been performed so that
similar transient events might have gone unnoticed. 
Recently, high-coherence
oscillations at frequencies between 5 and 8 Hz have been reported
from a number of Black-Hole Candidates. These are classified 
as 'type B' QPOs by Homan et al. (2001; see also Wijnands et al. 1999b
and Remillard et al. 2002).
In an observation of GX 339-4
during its 2002 outburst, a $\sim$6 Hz QPO was observed: it
appeared suddenly in the light curve and, once a random jitter in 
frequency with a characteristic time scale of $\sim$10s was removed, 
the corresponding PDS peak 
was very narrow (Nespoli et al. 2003). In this case however, the 
appearance of the oscillation was accompanied by significant flux, 
noise and spectral changes. Interestingly, in the interval just
before the sudden onset of the QPO, a broad and weaker (in terms
of fractional rms) peak was
present in the PDS, similar to what we found for 4U 1820-30. 
The observation where this QPO from GX 339-4 was seen corresponds
to the highest-rate interval of Very High State, 
possibly indicating the highest
accretion rate reached during the outburst. Notice that this time
also corresponded to the inferred ejection time for relativistic
jets in the system (see Fender \& Belloni 2004a,b).
A similar QPO with variable frequency 
between 5 and 8 Hz through different observations, was seen in the
bright transient XTE J1859+226 (Casella et al. 2004).
In that case, a number of sharp transitions were observed, including events
when the QPO appeared and disappeared within about $\sim$100 seconds. 
These type-B oscillations have well defined properties and seem
to appear in many sources (see also Takizawa et al. 1997, Homan et
al. 2001) when the accretion rate is high.
Notice that these QPOs are very distinct from the more common
'type-C' QPOs, the frequency of which 
is observed to vary between 0.1 and 20 Hz.
These are associated to strong band-limited noise and have a very
different phase-lag behaviour (see Wijnands et al. 1999b,
Casella et al. 2004).

At present, it cannot be ruled out that a similar phenomenon
is present in bright Z sources, at their highest inferred accretion rates, i.e.
in the flaring branch. The PDS analysis of the light curves of 
these sources is usually performed over time intervals
considerably longer than a few seconds, making it
possible that similar features were missed in past studies. This means
that a thorough search through the archive might reveal more cases, both
in atoll and Z systems.

It is tempting to associate the QPO we detected in 4U 1820-30 with
the type-B oscillations in BHCs, as they share many properties (note that due
to the short lifetime of the signal, we could not obtain an 
unambiguous determination
of the phase lags). 
The oscillation presented here are of course of much shorter duration than
anything seen so far in BHCs (see above), but what is strikingly similar is the
sharp appearance and disappearance of these features, not shared by any other
class of QPOs known to date.
The main problem is of course that 
one of the properties they share is the frequency range in which they
are observed, i.e. 4-7 Hz. Since BHCs of different (estimated) mass and
a NS system show about the same frequency, this means that if these
QPOs have the same origin, the
dependence of their frequency on the mass must be weak.
Of course, it is also possible that more parameters than the mass
of the compact object determine the QPO frequency, 
in which case there might be different dependences that compensate
each other. In any case,
the appearance of transient high-coherence QPOs in different
systems in their brightest state points towards a common
mechanism for these.

The physical origin of the oscillation reported here is not clear, in 
particular in view of the scarce observational information available. 
Its clear association with a period when a broader oscillation around similar
frequencies is observed allows us to suppose that whatever process was at work
in broadening the signal might have stopped its action for a few seconds,
leading to the detection of a much narrower feature. In a framework of a
`blob' model, where the oscillation is produced by the superposition of a number
of narrower oscillations caused by blobs of gas at different radii around the
compact object, this would correspond to the presence for a limited time of only
one of those blobs. This of course would not apply to the type-B oscillations in
BHCs, which are observed for a longer time (see Nespoli et al. 2003). 
In models like that of Psaltis \& Norman (2004), where a particular radius in
the accretion flow acts as a passband filter, this would correspond either to a
single `active' radius, or to a different mode of oscillation, but a more
quantitative estimate would be problematic. 
However, we know that QPOs are not coherent oscillations, which means that there
is a physical origin to their observed width. With the current instruments, we
are not able to ascertain the nature of this broadening (whether frequency,
phase or amplitude modulation), which would give important insight on the
physical processes producing the oscillation. In some case, like the one
presented here, a more coherent signal is observed for a limited time with
properties not compatible with simple random fluctuations. These cases provide
good tests for the development of theoretical models. 

\begin{acknowledgements}

This work was partially supported by MIUR under CO-FIN grant 2002027145.
We thank Luigi Stella for useful discussions and careful reading of the
manuscript.

\end{acknowledgements}

\end{document}